\documentstyle[12pt]{article}
\newfont{\frak}{eufm10 scaled\magstep1}
\newfont{\extra}{msbm10 scaled\magstep1}

\newcommand{\extr}[1]{\mbox{\extra #1}}
\newcommand{\sect}[1]{\setcounter{equation}{0}\section{#1}}
\newcommand{\subsect}[1]{\subsection{#1}}

\newcommand{\R}{\mbox{$I \!\! R \, $}}

\newcommand{\ba}{\begin{array}}
\newcommand{\ea}{\end{array}}
\newcommand{\be}{\begin{equation}}
\newcommand{\ee}{\end{equation}}
\begin{document}

\begin{center} 
{\bf O. Arratia, M. A. Mart\'{\i}n and  M. A. del Olmo} 
{\footnote{This work has been partially supported by a
DGES project (PB95--0719) from the Mi\-nis\-terio de Educaci\'on y
Ciencia de Espa\~na.  M.A.O. thanks to G. Marmo and A. Simoni for their
hospitality.    }}  
\end{center}
\vskip 0.5cm

\begin{center} {\LARGE{\bf{Moyal Quantization and group theory }}}
\end{center}  
\bigskip

\begin{center}
{\sl Dedicated to W.M. Tulczyjew in honor of his 65th birthday}
\end{center}  

\bigskip
\begin{abstract} 
We deduce a kernel that allows the Moyal quantization of the cylinder
(as phase space) by means of the Stratonovich--Weyl
correspondence.      
\end{abstract}

\sect{Introduction}

Moyal's work \cite{Moyal} showed how the Weyl correspondence
\cite{Weyl} may develop Quantum Mechanics (QM) as a theory of functions
on phase space. The algebra of these functions is endowed with a 
non-commutative product (star or Moyal product), while the physical 
states  are represented by the Wigner functions \cite{Wigner}. 

This formulation of QM has not had a success like those of Heisenberg, 
Schr\"odinger or Feynman. Difficulties like the extension of the theory
to particles with spin or to relativistic particles were some of the
reasons for this lack of success. In the last years, these
difficulties have been solved \cite{GVa}--\cite{CGV} by using a new
approximation to the problem: the Stratonovich--Weyl (SW)
correspondence. 

The seminal ideas of this procedure appeared many years ago in Ref.
\cite{Str}. The SW correspondence is applied to physical systems with a
(connected) Lie group of symmetries, $G$. The basic point is the SW 
kernel,  that maps the points of a coadjoint orbit (phase space) of $G$
into the set of operators acting on the carrier Hilbert space of the
projective unitary representation associated to this coadjoint orbit.    

However, all the cases studied in the above mentioned papers correspond
to free particles. The method has been applied in more general 
situations using central extensions of the symmetry group,
since in some cases, these extensions can be interpreted as
constant forces. Thus, some particular interactions
have been incorporated to the theory\cite{GMNO}--\cite{MO}. 
On the other hand, in recent works \cite{AOa}, \cite{AOb}, the problem 
of  contracting representations of groups has been studied for some 
kinematical groups and the results have been applied to the contraction 
of SW kernels. In physical terms, we have studied the contraction of 
classical systems together with the contraction of their quantized 
counterparts.

Nevertheless, not all the results have been satisfactory. In Ref.
\cite{GMNO} one of the groups under consideration was the euclidean
group of the plane, and the construction of the SW kernel for one of
the coadjoint orbits of this group (a cylinder) failed. Till now all
subsequent attempts have been useless. In this work we obtain
a SW kernel for that coadjoint orbit of the euclidean group by means of 
a constructive method. 

The paper is organized as follows. Section 2 is devoted to present two
short reviews about Moyal quantization and the SW correspondence. In
Section 3 we introduce the SW kernel for the cylinder and give the basic
hints for its construction. The paper ends with  some remarks and
conclusions.

\sect{Moyal quantization}

From the beginning of QM it has had a great interest on  reinterpreting 
the standard formulation in terms of Hilbert  spaces and operators using
the ``classical language" of phase spaces and functions defined on them,
trying to reduce it to a classical statistical theory.  Moyal in 1949
\cite{Moyal} took into account this philosophy and showed how to
construct such a theory for the case of euclidean phase space, i.e.,
$\R^{2n}$, making use of the Weyl correspondence \cite{Weyl} and the
Wigner functions \cite{Wigner} (see \cite{Gad} for a review). 

\subsect{Moyal formulation of Quantum Mechanics}

Moyal's formulation considers both  observables and states as 
(generalized) functions on a given phase space $M$, in such a way that 
we are able to  compute expected values by integrating over the phase
space as  in classical statistical mechanics
\begin{equation}
 \langle f \rangle_{\rho}= \frac{ \int_M f(u) \rho (u) du }{ \int_M
  \rho(u) du} 
\end{equation}

The process of quantization is introduced considering an associative
but non commutative ``twisted product''
\begin{equation}
  (f \times h)(u) = \int_M \int_M L(u,v,w) f(v) h(w) \, dv \, dw,
\end{equation}
where $L(u,v,w)$ is a non-local integral kernel to be defined. 

The problem of quantization is much more clarified when the system 
under study has a Lie group $G$ as set of symmetries. This group acts on the
phase space by symplectomorphisms, and induces a compatibility condition on 
the twisted product ($f^g \times h^g=(f \times h)^g$, with 
$f^g(u)=f(g^{-1}u)$) that is reflected on the trikernel $L$ in the
following covariance condition 
\begin{equation}
 L(g\cdot u,g\cdot v,g\cdot w)= L(u,v,w), \qquad \forall g\in G.
\end{equation}
In this case the (elementary) classical system can be identified 
with a coadjoint orbit of $G$ (or of a central extension of it) which 
will be  the phase space, and the quantum analog (in the standard
sense) with a projective unitary representation of G. The passage from
the classical to the quantum system is easily achieved with the aid of
the Kirillov theory~\cite{Kir} for the case of nilpotent groups, in
other cases the problem is more difficult.

A link between this geometric quantization and Moyal's one is provided 
by the SW correspondence which is explained in the next subsection.

\subsect{The Stratonovich--Weyl correspondence}

The SW correspondence contains, as a particular case, the Weyl
correspondence rule assigning linear operators on the standard Hilbert
space of QM to functions defined on a flat phase space. In this sense,
we can consider the SW correspondence  as a generalization of Weyl's
one, that allows us to extend Moyal QM to other situations (spin,
relativistic systems, systems under interaction, etc.). It is defined by
means of the SW kernel, ${\bf \Omega}$, through the integral formula 
\begin{equation}  
A= \int_{O} f({u}) 
{\bf \Omega}({u}) d\mu({u}).\label{ba} 
\end{equation}

The construction of ${\bf \Omega}$ exploits the fact that the
physical system under study has a (connected) Lie group $G$ as
symmetry group. The idea is to consider a projective unitary irreducible
representation (PUIR), $U$, of ${G}$ in a separable Hilbert space, 
${\cal H}$, and a coadjoint orbit, $O$, of ${G}$ (taken as phase space)
associated with it. The SW kernel is a mapping transforming each point
$u$ of $O$ into an operator, ${\bf \Omega}(u)$, acting  on ${\cal H}$
and verifying the following properties: 

\begin{enumerate}
            
\item $ {u} \longmapsto {\bf \Omega}({u}) $ is one to one.

\item $ {\bf \Omega}({u})$ is selfadjoint, $\forall u \in O$.

\item ${\rm tr} [ {\bf \Omega}({u})]=1,\;\; \forall {u} \in O$. 
This trace is usually defined in a generalized sense.

\item Traciality:   
\begin{equation}  
\int_{O} {\rm tr}[ {\bf \Omega}({u})  {\bf \Omega}(v) ]  
{\bf \Omega}( v) d\mu(v) =  {\bf \Omega}({u}),  \label{bb}     
\end{equation}
where  $\mu$ is the $G$--invariant measure on $O$. This property
means that ${\rm tr}[ {\bf \Omega}({u}) 
{\bf \Omega}(v)]$ behaves like a Dirac delta $\delta({u - v})$ 
with respect to the measure
$\mu(v)$.

\item Covariance:  
\begin{equation} 
U(g) {\bf \Omega}({u}) U(g^{-1})= 
{\bf \Omega}(g {u}),\quad \forall g \in {G}, 
\;\forall {u} \in O, \label{bc}
\end{equation}
with $gu$ the transformed point of $u$ by the coadjoint action of $g$.
\end{enumerate}

The construction of ${\bf \Omega}$ presents some problems. The first one
is related with the use of PUIR's: usually, it is easier to handle with
linear representations than with projective ones. This difficulty is
avoided by considering another group, $\overline{G}$, that linearizes 
the problem. The representation or splitting group $\overline{G}$ of $G$
is defined as the minimal  connected and simply connected central
extension of $G$ such that any PUIR of $G$ can be lifted to a linear
unitary irreducible representation (LUIR) of $\overline{G}$ and,
reciprocally, every LUIR of $\overline{G}$ provides  a PUIR of $G$.
 
The second problem appears as a consequence of the necessary 
relationship between coadjoint orbits and representations. They can be
considered as the classical and quantum version, respectively, of the
``same" physical elementary system, and therefore we should assign to
each LUIR of $\overline{G}$ a coadjoint orbit of $\overline{G}$. This
association is carried out by the method of Kirillov~\cite{Kir} for
constructing induced representations in the case of nilpotent groups.
For non nilpotent groups the association is not immediate, but can be
achieved in many cases.

The property of traciality permits to obtain an inversion formula for
the correspondence
\begin{equation} 
W_A({u})\equiv {\rm tr} [A {\bf \Omega}({u})]= \int_{O} 
 f({v}) {\rm tr} [ {\bf \Omega}({u}) {\bf \Omega}(v)] 
 d\mu({v})= f(u). \label{bd}
\end{equation}
The function $W_A({u})$ is usually called the Wigner function of $A$,
with reference to the functions obtained by inverting the Weyl mapping,
as Moyal  pointed out.
Traciality also yields the following expression
\begin{equation} {\rm tr} [AB] = \int_{O} W_A({u}) 
W_B({u})  d \mu({u}), \label{be}
\end{equation}
allowing to obtain quantum averages as in classical statistical 
mechanics, the centerpiece of Moyal's formulation.

  Physical calculations based on the SW correspondence are made by means 
of a noncommutative product, the so-called star or twisted product, for 
generalized functions on phase space. This is equivalent to the product
of operators on its corresponding Hilbert space. We can define the
twisted product of two functions $f({u})$ and $g({u})$ on $O$ as
\begin{equation}  
(f*g)({u})=\int_{O}\int_{O} {\rm tr}
[{\bf \Omega}({u}){\bf \Omega}(v){\bf \Omega}(w)]\;f(v)
g(w)d\mu(v)d\mu(w). \label{bf}
\end{equation}
It is easy to verify 
\be 
(W_A*W_B)({u})=W_{AB}({u}), \label{bg} 
\ee
and 
\be \int_{O}(f*g)({u})d\mu({u})\;=\;\int_{O}f({u}) g({u})d\mu({u}).
\label{bh}
\ee 
The term ${\rm tr}[{\bf \Omega}(u){\bf \Omega}(v){\bf \Omega}(w)]$ is
called the tri-kernel of the SW correspondence and is the main
ingredient for quantization, together with the coadjoint orbit (phase
space).
 
There is no canonical way of constructing SW kernels and no theoretical
result is known assuring the existence or unicity of these kernels. 
However, for many physical systems the problem has been solved 
(Ref. \cite{CGV}, \cite{GMNO}) following three simple steps: 1) Choose
an arbitrary point ${u}_0$ of $O$ as origin. 2) Make an Ansatz  for a
selfadjoint operator, ${\bf \Omega}({u}_0)$, of trace one (with respect
to a suitable trace). 3) Finally, define the kernel on the whole  $O$ by 
\begin{equation}  
{\bf \Omega}( {u})= {\bf \Omega}(g {u}_0) 
 = U(g) {\bf \Omega} ({u}_0) U(g^{-1}), \label{bi}
\end{equation}
where  $g$ is an element of $\overline{G}$ such that $g {u}_0 = {u}$.  
Note that this kernel is well defined if and only if
  \begin{equation} 
{\bf \Omega}({u}_0) = U(\gamma) {\bf \Omega} 
 ({u}_0) U(\gamma^{-1}),\;\;\; \forall \gamma \in \Gamma_{{u_0}},  
\label{bj}
\end{equation}
where $\Gamma_{{u}_0}$ is the isotopy group of ${u}_0$, i.e., 
$\Gamma_{{u}_0}=\{\gamma \in \overline{G}\  |\ \gamma {u_0}={u_0}\}$.
This property, proved in Ref.~\cite{GMNO}, implies that ${\bf \Omega}
({u})$,  defined as above, is covariant. Reciprocally, if ${\bf
\Omega}({u})$ is covariant, the latter property holds. Remark that the
covariance property guarantees that  the SW kernel is well defined on
the coadjoint orbit $O$, in other words, it is independent of the choice
of a section from  $O$ on $\overline{G}$.
\medskip

{\bf Example 2.1.-} We can give an interesting and simple example to
improve the understanding of the previous construction for the standard
quantum theory. 

 The basic tool in the Moyal formulation of QM~\cite{Gro},~\cite{Litle} 
is the twisted product (also Moyal product) for functions on phase
space.  This product can be defined by using the Weyl mapping, i.e., a
linear  isomorphism between the space of the above mentioned functions
and the  space  of the operators on a standard Hilbert space. The Weyl
mapping can  be introduced  through the Grossmann--Royer 
operators~\cite{Gros},~\cite{Roy}, which are defined as follows 
\be 
[{\bf K}({\bf q},{\bf p})\varphi]({\bf x})=2^n
e^{2i{\bf p}\cdot({\bf x}-{\bf q})}  \,\varphi(2{\bf q}-{\bf x}),
\label{bk}
\ee
where the standard $\R^{2n}$ phase space with canonical coordinates 
$({\bf q},{\bf p})$ is assumed. These operators act as integral kernels 
in such a way that to a function $f$ corresponds the operator
\be  
W(f)= \frac{1}{2\pi}\int_{\R^{2n}} f({\bf q},{\bf p}) 
{\bf K}({\bf q},{\bf p}) d{\bf q}\,d{\bf p}. \label{bl}
\ee
The mapping is invertible, so the Moyal product can be defined by
\be 
f*g=W^{-1}(W(f)W(g)),\label{bm}
\ee
whose explicit expression is
\be  
(f*g)({\bf u})=\frac{1}{\pi}\int_{\R^{4n}} f({\bf v}) 
       g({\bf w})\exp[i({\bf u}J{\bf v}+{\bf v}J{\bf w}
         +{\bf w}J{\bf u})] d{\bf v}\,d{\bf w}, \label{bn}
\ee
where $J$ is the matrix $\left(\ba{cc}0 & I_n\\-I_n & 0 \ea\right)$,
$I_n$ is the $n\times n$ identity, and ${\bf u},{\bf v}$ and ${\bf w}$
stand for $({\bf q},{\bf p}),\, ({\bf q}',{\bf p}')$ and
$({\bf q}'',{\bf p}'')$, respectively.

Now, we can construct the SW correspondence for the Heisenberg 
group $H^{2n+1}$, i.e., the set $\R^{2n+1}$ endowed with the following
composition law 
\be 
({\bf a},{\bf b},c)({\bf a}',{\bf b}',c')=
({\bf a}+{\bf a}',{\bf b}+{\bf b}',c+c'+\frac{1}{2}          
({\bf a}\cdot{\bf b}'-{\bf a}'\cdot{\bf b})),\label{bp}
\ee
with ${\bf a},{\bf a}',{\bf b},{\bf b}'\,\in\R^n$ and $c,c'\,\in\R$.

The corresponding Lie algebra, ${\cal H}^{2n+1}$, is a
$(2n+1)$--dimensional one, generated by $I$, ${\bf Q}$ and ${\bf P}$.
In the standard formulation of QM these generators are represented as
the identity, position and momentum operators on $L^2(\R^n)$,
respectively,  with nonvanishing commutation relations
$[Q_i,P_j]=iI\delta_{ij}$. 

The coadjoint action is given by 
\be
{\bf x} ' = {\bf x} + z{\bf b}, \quad {\bf y} '= {\bf y} - z{\bf a},
\quad z '=z , \label{bpa}
\ee
where 
$({\bf x},{\bf y},z)$ are the coordinates of a point of $({\cal
H}^{2n+1})^*$ in a basis dual of the basis $\{Q_i, P_i, I \}$.
       
The coadjoint orbits of dimension greater than zero are all of them
isomorphic to $\R^{2n}$, and yield the same kind of induced
representations. A LUIR of $H^{2n+1}$ associated to the coadjoint
orbit specified by $z=1$ is 
\be  
[U({\bf a},{\bf b},c)\varphi]({\bf \xi})=\exp
[-i(c+{\bf b}\cdot{\bf \xi}
 +\frac{1}{2}{\bf a}\cdot{\bf b})]\,\varphi({\bf a}+{\bf \xi}), 
\label{bq}
\ee
where $\varphi \,\in\,L^2(\R^n)$.

Taking $({\bf q}={\bf x}/z,\,{\bf p}={\bf y})$ as canonical
coordinates on the orbit and choosing the point $({\bf 0},{\bf 0})$ as
origin $(u_0)$ we can make the Ansatz
\be 
[{\bf \Omega}(u_0)\varphi]({\bf \xi})=2^n \,\varphi(-{\bf \xi}),
\label{br}
\ee
and obtain from it the SW kernel and the trikernel, which coincide with
(\ref{bk}) and the integral kernel in (\ref{bn}), respectively.

\sect{Stratonovich--Weyl kernel for the cylinder}

In this section we will finally accomplish the quantization of the 
cylinder (considered as a coadjoint orbit of the euclidean group) by
means of the construction of an appropriate SW kernel. We start
presenting some basic notions about the euclidean group that will be
used later. A brief review about the preceding attempt \cite{GMNO} for
quantizing the cylinder is also included for the sake of completeness.

\subsect{The Euclidean group of the plane}

The euclidean group $E(2)$ is the group of motions of the plane,
i.e., the set of transformations leaving invariant the euclidean length
in $\extr{R}^2$, roughly speaking, translations and rotations. 

The corresponding Lie algebra, ${e}(2)$, has dimension three and is
spanned by the infinitesimal generators of translations, $P_1$ and
$P_2$, and rotations around an axis perpendicular to the plane, $J$.
Their commutation relations are   
\be
[J, P_i]= \epsilon_{ij} P_j, \quad  [P_1,P_2]= 0. \quad i,j=1,2, 
\label{ca}
\ee
where $\epsilon_{ij}$ is the totally skewsymmetric tensor.
The algebra ${e}(2)$ admits a maximal nontrivial central 
extension by \extr{R}, associated to a new non-zero Lie bracket:
$[P_1,P_2]= I$, where $I$ is the new central generator. 

Integrating the preceding commutators we get the group law for the
extended euclidean group $\bar{E}(2)$, 
\be 
g'g=(\eta', {\bf a'}, \phi')(\eta, {\bf a}, \phi)=
(\eta'+\eta+ \frac{1}{2} {\bf a}^{\phi'} \times {\bf a}, 
{\bf a'} + {\bf a}^{\phi'}, \phi'+\phi), \label{cb}
\ee
where $g\equiv(\eta, {\bf a}, \phi)= e^{\eta I} e^{\bf a P} e^{\phi J}$
($\eta \in \R,\ {\bf a} \in \R ^2,\ \phi \in [0,2\pi)$), and ${\bf
a}^{\phi}$ stands for the transformed of the vector ${\bf a}$  by a
rotation of angle ${\phi}$. 

The coadjoint action of $\bar{E}(2)$ on $\bar{e}(2)^*$, the dual of
$\bar{e}(2)$, can be readily computed. Its expression, using coordinates
$(\beta, {\bf p}, j)$ in the dual basis to  $(I, {\bf P}, J)$ is 
\be 
\beta'    =  \beta, \qquad
{\bf p'}  =   {\bf p}^\phi + \beta {\bf a}^\frac{\pi}{2}  , \qquad
j'=   j + {\bf a} \times {\bf p}^{\phi} +  \frac{1}{2} \beta {\bf a}^2 .
\label{cc}
\ee
This action splits $\bar{e}(2)^*$ into several orbits, which can be
classified according to the invariants $\beta$  and 
${\bf p}^2- 2 \beta j$ as follows: 
\begin{enumerate}
\renewcommand{\labelenumi}{(\roman{enumi})}
\item $\beta \not = 0$,  this kind of orbits are 2D paraboloids.
\item $\beta =0$ and ${\bf p}^2 = r^2 \not = 0$, these are 2D cylinders
parallel to the $j$-axis, which will be denoted by ${\cal O}_r$.
\item $\beta =0 $ and ${\bf p}^2 = 0$, each point $(0,{\bf 0},j)$ is a 
zero dimensional orbit.  
\end{enumerate}

Orbits of type (i) were studied in \cite{GMNO} where their complete
Moyal quantization program was accomplished. However, for the second
kind of them the results were not completely satisfactory because the
proposed kernel did not verify traciality. In subsect. 3.3 we will
present a suitable SW kernel for these orbits. The third kind of orbits
are meaningless from the physical point of view because they represent
systems without dynamics.

\subsect{Previous results about cylinder quantization}

The $\bar{E}(2)$--homogeneous space  can be endowed naturally with a
symplectic structure. A set of canonical coordinates for this
structure is $\alpha= {\rm Arg} ({\bf p})$ and $j$, i.e., $ \{\alpha,
j\}=1$ as is  easy to check.

The representation $U_r$ associated to ${\cal O}_r$ is obtained by
Kirillov's theory \cite{Kir}
\be 
[U_r(\eta, {\bf a}, \phi)\psi](\theta)= 
e^{i {\bf a} \cdot {\bf t} } \psi(\theta-\phi), \label{da} 
\ee
where ${\bf t}= r(\cos \theta, \sin \theta)$ and $\psi$ belongs to
$L^2(S^1)$.

In order to construct the kernel ${\bf \Omega}$ we take into
account that in other physical situations \cite{GMNO} parity-like
operators (\ref{br}) have played a remarkable role in the definition of 
the SW kernel at the origin point of the orbit. Hence, it is reasonable
to consider in this case the operator 
\be 
[\Omega(0,0)\psi](\theta)= N \psi(-\theta), \label{db}
\ee
where $(0,0)$ are the canonical coordinates of the chosen origin 
point, and $N$ is an arbitrary normalization constant.
This operator verifies property (\ref{bc}) and, henceforth, it induces
a well defined object on the whole orbit:
\be 
\mbox{[}\Omega(\alpha,j) \psi](\theta)= N 
 e^{2ij \sin(\alpha- \theta)} \psi(2 \alpha- \theta). \label{dc}
\ee
Operators just calculated are self-adjoint and of unit trace when $N=2$.
In addition, they satisfy the covariance property but, unfortunately,
they do not satisfy traciality, i.e., we have  
\be 
{\rm tr}[ \Omega(\alpha',j') \Omega(\alpha, j)]=
4 \pi \delta(\alpha-\alpha') J_0(2(j-j')), \label{dd}
\ee
with $J_0$ the first-kind Bessel function. 

Other parity-like operators as 
\be
[\Omega(0,0) \psi](\theta)= N \psi(\theta + \pi), \quad {\rm or}
   \quad [\Omega(0,0) \psi](\theta)= N \psi(-\theta + \pi) \label{de}
\ee 
lead to loose even the covariance.

\subsect{SW kernel for the cylinder}

A new approach to this subject consists in taking an arbitrary 
operator  at the origin point of the orbit and then to impose
the conditions 1)--5) to it. The most restrictive conditions are
covariance and traciality. Next proposition is the result of requiring
covariance. 

\medskip

{\bf Proposition 3.1.-}{\sl The following statements are equivalents:

{\rm \ \ (i).} $\Omega$ is covariant. 

{\rm  \ (ii).} $U(\gamma) \Omega(u_0)
U(\gamma^{-1})= \Omega(u_0), \qquad \forall \gamma \in \Gamma_{u_0}$. 

{\rm (iii).} $[U(X), \Omega(u_0)]=0,  \qquad \forall X \in
{Lie}(\Gamma_{u_0})$.} 

{\sl Proof}.- The equivalence of (i) and (ii) can be
find in Ref. \cite{GMNO}. We leave the rest of the demonstration to the
reader. 

For the case we are interested in, the isotopy group $\Gamma_{u_0}$ is
generated by $P_1$. The space $L^2(S^1)$ admits a discrete basis given
by the functions $\{f_n(\theta)= \frac{1}{\sqrt{2 \pi}} e^{in \theta};\
n \in \extr{Z}\}$ which, hereafter, will be written in the bracket
notation  $ \{ |n \rangle ;\ n \in \extr{Z}\}$. 

Substituting $\Omega(u_0)$ by a generic operator $A$ in 
statement (iii) of Prop. 4.1, and computing  $\langle r | [P_1, A]
| s \rangle$ we obtain 
\be
\langle r | [P_1, A] | s \rangle = 
 A_{s, r+1} + A_{s, r-1} - A_{s+1, r}- A_{s-1, r} = 0.  \label{eb}
\ee
 Using techniques from the theory of finite difference equations
we find the general solution of the preceding system
\be 
A_{r,s}= a_{r+s} + b_{r-s}, \label{ec}
\ee
where $a$ and $b$ are arbitrary functions on $\extr{Z}$, that can also
be considered as Fourier coefficients of two functions $a$ and $b$ on 
$S^1$.

The explicit action of $A$ on a generic function $f \in L^2(S^1)$ is  
\be 
[Af](\theta)= a(\theta) f(- \theta) + b(\theta) f(\theta).
\label{ed}
\ee 
Thus, we obtain the more general covariant operator-valued function on
the cylinder  
\be
[{\bf\Omega}(\alpha, j)f](\theta)= e^{2ij \sin(\theta-
\alpha)} a(\theta-\alpha) f(2 \alpha- \theta) + b(\theta- \alpha) 
f(\theta). \label{ee}  
\ee
Imposing the hermiticity property to this operator we get that the
function $b$ must be zero and that $a$ verifies the relation
$a(-\theta)=  \overline{a(\theta)}$, where the bar means complex
conjugation.

The SW correspondence associated to (\ref{ee}), once made $b=0$, maps
the set of operators on the Hilbert space $L^2(S^1)$ into the set of 
functions on the cylinder. These operators can be written in terms
of the transition operators $P_{n,m}= |m\rangle \langle n|$. The 
associated symbol to $P_{n,m}$ is a function that can be factorized in
the following way 
\be 
W_{n,m}(\alpha, j)=
\Theta_{m-n}(\alpha) L_{m+n }(j),\label{ef}  
\ee 
with 
\be
\Theta_n(\alpha)= \frac{1}{\sqrt{2 \pi}} e^{i n \alpha}, 
\qquad L_n(j) = \int_{S^1} \frac{d\theta}{2 \pi} 
e^{2ij \sin \theta} a(\theta) e^{-in \theta}.\label{eg} 
\ee
Note that if we choose $a(\theta)=1$, then we reproduce
the case given by (\ref{dc}) and the term (\ref{eg}b)
reduces essentially to a Bessel function, i.e.,  $L_n(j)= J_n(2j)$.

Traciality property (\ref{bb}) can be rewritten, more conveniently
for our purposes, in terms of the functions $W_{n,m}$ as
\be 
\int_{\cal O} d\mu(v) K(u,v) W_{n,m}(v) = W_{n,m}(u),\label{eh} 
\ee
where  $K(u,v) = {\rm tr}[\Omega(u) \Omega(v)]$. This last expression
allows to interpret $K$ as a reproducing kernel on  the space spanned by
the functions $W_{n,m}$. Note that $K(u,v)= \sum_{l,k} W_{l,k}(u)
W_{k,l}(v)$. Using covariance and the factorization (\ref{ef}),
condition (\ref{eh}) can be reduced to  
\be 
\sum_s L_{2s+r}(u_0) \langle L_{2s+r}| L_r \rangle = L_r(u_0). 
\label{ei}
\ee
Finally, a sufficient condition in order to get traciality is to impose
the following condition on $a$: \be 
| a(\theta) |^2 + | a(\theta+\pi) |^2 = 4 |\cos \theta|.\label{ej}
\ee

Summarizing, we can say that the operators defined by
\be 
[ \Omega(\alpha, j) \psi](\theta) = e^{2ij \sin (\theta-\alpha)}
a(\theta-\alpha) \psi(2\alpha- \theta), \label{ek}
\ee
with $a$ satisfying $a(-\theta)= \overline{a(\theta)}$ and (\ref{ej}),
are covariant, self-adjoint and tracial.

The following two remarks concern the requirements about trace, 3), and
injectivity, 1), that any SW kernel ought to verify:

$i$) The trace of the operator $\Omega(\alpha, j)$ is equal to 
$\frac{1}{2} a(0)$, so we must take $a(0) \not = 0$ in order to get a
finite trace. 

$ii$) Function $a$ can be written as 
\be
a(\theta) = 2 \sqrt{ | \cos \theta | } 
\cos( \frac{\pi}{4}+ h(\theta) ) e ^{i \varphi(\theta)},\label{el} 
\ee
where the functions $h:S^1 \longrightarrow
[-\frac{\pi}{4},\frac{\pi}{4}]$ and  $\varphi:S^1 \longrightarrow
(-\pi,\pi]$ satisfy the conditions $h(\pi +\theta) = - h(\theta)$ and 
$\varphi(-\theta)= -\varphi(\theta)$. When $h \not = 0$ the SW kernel is
injective. In  the case $h=0$ we still have two possibilities depending
on the value of  $ \varphi(\theta + \pi) -\varphi(\theta)$: it can be 
different from or equal to $2 \sin\theta$. The kernel $\Omega$ is 
injective in the first situation, but is not in the second one.

\sect{Conclusions}

The problem of the cylinder quantization has been solved by the SW
procedure once the SW kernel was known. The constructive nature of the 
method, used here for obtaining the SW kernel, allows to get the
general solution imposing successively the appropriate conditions to
have suitable kernels, without making any Ansatz about the value of the
kernel at a point. Our deduction of the SW kernel is strongly  based on
Prop. 3.1 by taking advantage of the infinitesimal version for
covariance property established in it.   

On the other hand, it is worthy to note that we have found not a single
kernel but a whole family of SW kernels depending on a function. An open
problem to investigate is whether different kernels give different
quantizations or not. Another researching direction is to impose new
conditions in order to guarantee the unicity of the kernel. Work along
both lines is in progress.


\bigskip

Oscar ARRATIA  (e. mail: oscarr$@$wmatem.eis.uva.es)

 Miguel A. MART\'IN (e. mail:migmar$@$wmatem.eis.uva.es)

Departamento de Matem\'atica Aplicada a la Ingenier\'{\i}a,  

Universidad de Valladolid. E-47011, Valladolid, Spain
 
\medskip
Mariano A. del OLMO  (e. mail: olmo$@$cpd.uva.es)  

Departamento de F\'{\i}sica Te\'orica,

Universidad de Valladolid. E-47011, Valladolid, Spain


\begin{thebibliography}{99}

\bibitem{Moyal}  Moyal J.E., Proc. Camb. Phil. Soc. {\bf 45}, (1949), 99.

\bibitem{Weyl}  Weyl H., {\sl The Theory of Groups and Quantum
Mechanics}. Dover, New York, (1931).

\bibitem{Wigner} Wigner E.P., Phys. Rev. {\bf 40}, (1932), 749.

\bibitem{GVa} Gracia Bond\'{\i}a J.M.,  Varilly J.C., J. Phys. A: Math.
Gen. {\bf 21}, (1988), L879.

\bibitem{GVb}  Gracia Bond\'{\i}a J.M., Varilly J.C, Ann. Phys. (N.Y.) 
{\bf 190} (1989), 107.

\bibitem{CGV}  Cari\~nena J.F., Gracia Bond\'{\i}a J.M.,  Varilly J.C., 
J. Phys. A: Math. Gen. {\bf 23}, (1990), 901.

\bibitem{Str}  Stratonovich R.L., Sov. Phys. JETP {\bf 31}, (1956),
1012.

\bibitem{GMNO}  Gadella M., Mart\'{\i}n M.A., Nieto L.M., del Olmo
M.A.,  J. Math. Phys. {\bf 32}, (1991), 1182.
        
\bibitem{BGO}  Ballesteros A.,  Gadella M.,  del Olmo M. A.,
J. Math. Phys. {\bf 33}, (1992), 3379. 
        
\bibitem{MO}  Mart\'{\i}n M.A.,  del Olmo M. A., J. Phys. A: Math. Gen.
{\bf 29}, (1996), 689.
        
\bibitem{AOa}  Arratia O.,  del Olmo M.A.,
{\sl Elementary systems of $(1+1)$ kinematical groups: contractions and
quantization}. To appear in Fortschr. Phys.
 
\bibitem{AOb}  Arratia O., del Olmo M.A.,
{\sl Contraction of representations of $(1+1)$ kinematical groups and
quantization}. To appear in Int. J. Mod. Phys. A.
 
\bibitem{Gad}  Gadella M., Fortschr. Phys. {\bf 43}, (1995), 229.

\bibitem{Kir}  Kirillov A. A., {\em Elements of the Theory of
Representations}. Springer, Berlin (1976).

\bibitem{Gro}  de Groot S. R., {\sl La transformation de Weyl et la 
fonction  de Wigner: une forme alternative de la m\'ecanique quantique}..
Les Presses de l'Universit\'e de Montr\'eal, Montr\'eal (1974).

\bibitem{Litle}  Littlejohn R.G., Phys. Rep. {\bf 138}, (1986), 193.

\bibitem{Gros} Grossmann A.,  Commun. Math. Phys. {\bf 48}, (1976), 191.

\bibitem{Roy}  Royer A., Phys. Rev. A {\bf 15}, (1977), 449.

\end{thebibliography}
\end{document}